\documentclass[conference]{IEEEtran}
\IEEEoverridecommandlockouts
\usepackage{cite}
\usepackage{amsmath,amssymb,amsfonts}
\usepackage{amsthm}
\usepackage{algorithmicx}
\usepackage[ruled]{algorithm}
\usepackage{algpseudocode} 
\usepackage{graphicx}
\usepackage{textcomp}
\usepackage{xcolor}
\usepackage[utf8]{inputenc}
\usepackage[english]{babel}

\usepackage{multirow}
\usepackage{makecell}
\usepackage{hhline}
\usepackage{caption}
\usepackage{subcaption}
\usepackage{color}
\def\BibTeX{{\rm B\kern-.05em{\sc i\kern-.025em b}\kern-.08em
    T\kern-.1667em\lower.7ex\hbox{E}\kern-.125emX}}

\begin{document}

\title{Machine Learning Enhanced Blockchain Consensus with Transaction Prioritization for Smart Cities
}
\author{S. Valli Sanghami, John J. Lee, and Qin Hu  (Corresponding Author)\\ % <-this % stops a space
Indiana University-Purdue University Indianapolis \\
Email: \{vshanka,johnlee,qinhu\}@iu.edu
}

\maketitle

\begin{abstract}
In the given technology-driven era, smart cities are the next frontier of technology, aiming at improving the quality of people’s lives. Many research works focus on future smart cities with holistic approach towards smart city development and realising an overarching smart city vision.
In this paper, we introduce such future smart cities that leverage completely on blockchain technology in areas like data security, energy and waste management, urban management, governance, transport, supply chain, including emergency events and environmental monitoring. Blockchain, being a decentralized immutable ledger, has the potential to promote the development of smart cities by guaranteeing
transparency, data security, reliability, efficiency, interoperability and privacy which makes it a promising fit for smart cities.  Particularly, using blockchain in emergency events will ensure multiple parties to coordinate resources in an emergency, coordinate more efficient disaster responses, enable rescue teams to perform their jobs more efficiently, provide interoperability between many parties involved in response, increase timeliness of services and provide transparency. In that case, if a current fee-based or first-come-first-serve-based processing is used, emergency events may get delayed in being processed due to competition, and thus, threatening people's lives. Thus, there is a need for transaction prioritization based on the priority of information and quick creation of blocks (i.e., variable interval block creation mechanism). Also, since the leader will ensure transaction prioritization while generating blocks, leader rotation and proper election procedure becomes important for the transaction prioritization process to take place honestly and efficiently in our consortium blockchain.
In our consensus protocol, we deploy a machine learning (ML)  algorithm  to achieve efficient leader election and design a novel dynamic block creation algorithm. Also, to ensure honest assessment from the followers on the leaders’ generated blocks, a peer-prediction-based verification mechanism is proposed. Both security analysis and simulation experiments are carried out to demonstrate the robustness, accuracy, and efficiency of our proposed scheme.

\end{abstract}


\section{Introduction}
Digital technology has transformed every aspect of human life, enhancing our productivity and efficiency. Leveraging on various technology breakthroughs, smart cities become one of the first steps towards an intelligent future. Cities are beginning to evolve by integrating technological dynamism in operations of infrastructure maintenance, identity management \cite{22}, healthcare \cite{25}, transportation \cite{21}, energy distribution \cite{23}, and waste management, creating a sustainable living environment.

However, there are a few challenges that need to be overcome for smart cities deployment, such as sensor data integration, storage, security and privacy issues. 
%and henceforth such data need to be properly secured. Also, data will be shared among devices, and hence any manipulation or leakage of data should be avoided. 
For sensor data integration, numerous sensors are employed to collect necessary data in a timely manner, where the number of sensors will keep increasing for both comprehensiveness and accuracy considerations. %, and so integration of all such sensors and devices participating becomes far more critical. 
Data storage becomes an important challenge as the growing number of sensors will result in huge amount of generated data. Improper storage of those data can lead to inefficient management or even fatal errors.
%This big data plays a critical role in smart city-based applications as this data gives a better insight of overall operations in the city. Hence, data needs to be properly stored, as any improper storage may lead to the loss of critical data. Therefore, finding a better and faster storage infrastructure which is cost-effective and reliable becomes significant.
Data security is one of the significant aspects of smart cities since the smartness is practically dependent on data collected from various devices, ranging from sensors for rush hour statistics to air quality measurements, and data sharing among them for critical decision making. Any malicious manipulation or unexpected leakage of smart-city data can bring huge economic loss or even threaten people's lives. 

Surveillance capitalism is also another threat to individual privacy. It would result in distrust and skepticism among different parties in sharing their data, owing to the fear of any potential compromise of the data. Digital information is more valuable  than  physical  currencies  in  smart cities running as a  digital world.  Hence, any weak spot in the system will cause the system to  become a target of hackers and will cause data breach. This will impact the entire system, enabling malicious actors to obtain critical information or render vital information useless. Considering an autonomous vehicle as an example,  the autonomous vehicle can be hacked and compromised by the hackers which can lead to the vehicle's brake/acceleration being controlled by the attackers or the vehicle can be diverted on an unknown pathway by the hackers; or  these type of attacks can also result in issuing distrustful messages to the network and resulting in miscommunication between vehicles, thus causing severe accidents.
 
This is where  blockchain can come into play with its decentralized peer-to-peer (P2P) network topology and the shared distributed ledger, promising security and integrity of data, which thus  revolutionizes the development of smart cities. In particular, blockchain can be used to interconnect and integrate various departments, thus enabling different components in smart cities to well coordinate via storing critical information in a secure  and transparent manner \cite{27}. Besides, the employment of blockchain in smart cities can also facilitate smooth communication between the government sectors and citizens, thereby increasing efficiency and establishing trust. Also, blockchain can help engage community stakeholders in various decision-making cases with specifically designed procedures \cite{28}. In general, blockchain %is one of the most cost-effective solutions (if not PoW), as it 
removes intermediaries from the traditional smart-city systems, thus improving efficiency of the whole network.

There exist various existing research on applications of blockchain in smart cities, such as  maintenance of ID cards \cite{22} and health records \cite{25}, verifying and sharing legal documents, automotive industry \cite{21}, interoperability of smart devices \cite{26}, and energy distribution \cite{23}. %Many research works [1-6] investigate on deploying the blockchain network to various aforementioned application scenarios.
Considering that the blockchain network for smart city is rich with data, it becomes a necessity to categorize and prioritize those data as different urgency levels of transactions to be recorded on the blockchain. 
This prioritization feature will enable that any transaction carrying critical information, such as optimizing traffic lights, information about emergency vehicles, and notification regarding accidents, can be recorded on the main chain for timely response without any delay. However,
most of the existing works \cite{27,28,22,25,21,23,26} fail to consider about prioritizing transactions for a better and efficient governance. 
%Also, other consensus mechanisms focusing on scalability and efficiency improvement [7-9] are still in developing stages and face various security threats which have not been investigated thoroughly.

Transaction prioritization puts general welfare of the public ahead of the interests of individual blockchain nodes, such as obtaining higher transaction fees, and thereby aiding in creating a better world in both digital and physical senses. In order to ensure that the blockchain nodes truthfully implement transaction prioritization instead of concerning more about their self and instant interests at first,
we propose a novel and efficient consensus protocol for the permissioned blockchain system benefiting the governance of smart cities. %Particularly, data entering the network is efficiently managed by introducing transaction prioritization in the blockchain network.
All steps involved in our consensus mechanism are explained in detail based on algorithm designs. 
%Specifically, to choose efficient leaders for recording transactions in the expected manner, a machine learning algorithm and a random function based on entropy are jointly utilized. For efficient block generation, dynamic block creation policy is proposed. To motivate the blockchain nodes to submit honest feedback on the leader's work, peer-prediction mechanism with proper quadratic scoring rule is employed. 

In summary, main contributions of this work are listed as follows:
\begin{enumerate}
    \item A new consensus protocol supporting transaction prioritization in blockchain is designed to benefit smart city management efficiency, especially handling emergencies. 
    \item Transaction prioritization is implemented based on their urgency levels to record priority transactions for promptly response without any delay.
    \item To improve the trustworthiness and efficiency in the leader election process, we design a new algorithm jointly employing a machine learning (ML) algorithm, namely LightGBM, and a modified true random number generator (m-TRNG) function.
    \item A dynamic block creation algorithm is proposed  based on the presence of priority transactions in the network for achieving the trade-off between block generation cost and system efficiency.
    \item To motivate blockchain nodes to submit honest feedback on the leader's work, we put forward a peer-prediction-based mechanism using the quadratic scoring rule.
\end{enumerate}

The rest of this paper is organized as follows. Section II  investigates  the  most  related  work.  Our system model and its necessity are introduced in Section III. Section IV introduces the overall design of our proposed consensus protocol, where detailed steps are presented with algorithm designs. Section V analyzes the performance of our blockchain system against various common security attacks. Section VI displays the experimental evaluation results, and Section VII concludes the whole paper.

\section{Related Work}
% \subsection{Blockchain for Smart Cities}
Recent research works investigate different application scenarios of blockchain systems \cite{4,5}, such as smart cities, data management, healthcare, and financial sector, to reform and improve traditional practices. One such interesting area of application is smart cities which has garnered more attention from researchers  \cite{27,28,21,22,25,26,6,7,9,8,14}.

In \cite{27}, Hakak \text{et al.} analyzed the merits and challenges of adopting blockchain technology in smart cities, and presented blockchain-based architecture for smart cities, along with case studies. 
In \cite{28}, Scekic  \text{et al.} proposed a blockchain-based platform to 
facilitate ad-hoc interactions and encourage co-creation processes among citizens through a reward system.

Some research \cite{22,25,21,26} discussed the usage of blockchain technology in fundamental and critical applications like  identity management, healthcare, automotive industry, and interoperability of devices. In \cite{22}, Asamoah \text{et al.} examined the purpose of blockchain in identity management by the usage of verifiable attributes of users participating in the network. Meanwhile, Linn \text{et al.} \cite{25} focused on interoperability in healthcare by using  blockchain as an    access-control manager to manage and control access to patient's critical data. Here, various security attacks were not discussed. Sharma  \text{et al.} \cite{21} proposed a blockchain-based framework for supply chain management, maintenance, and recycling of automotive parts, and determined efficient nodes to avoid the mining process in PoW. However, they fail to discuss about efficiently managing the network traffic caused by big data from smart cities.
%Others \cite{25,26} discussed about achieving interoperability by using blockchain technology. 
Viriyasitavat \text{et al.} \cite{26} proposed a new architecture for interoperability of services by integrating blockchain with service-oriented architecture and key performance indicators to mitigate inconsistency in data formats and system interfaces, and to provide a second layer of security.

Others \cite{6,7,8,9} used  blockchain systems in smart city applications to solve various data-related issues and to create a  trustful environment. In \cite{6}, Biswas \text{et al.} predominantly focused on data storage and data management, while Ibba \text{et al.} \cite{7} and Sun \text{et al.} \cite{9} discussed more about improving data security and privacy protection by using blockchain-based services in smart cities.
Here, swiftly recording critical information entering the blockchain network is not clearly discussed. 
Further, Michelin \text{et al.} \cite{8} proposed a way to improve the efficiency of employing the blockchain network in smart cities by  decoupling the data stored in transactions from the block header. This allows for a faster recording of transactions. 

Pournaras \cite{14} proposed a consensus protocol for smart city-based applications, named Proof of Witness Presence, which engages citizens in decision-making of the governance. Here, transaction costs and latency in the network might affect the performance of this consensus protocol. 
Moreover, the above work did not investigate the performance of the proposed consensus protocol against various security attacks.

% \subsection{Consensus Protocols for Blockchain Systems}
%There are many research works \cite{11,12} focusing on designing new consensus protocols for different application scenarios with the aim of improving scalability and efficiency compared to the traditional protocols. 
%Milutinovic \textit{et al.} \cite{11} introduced randomness in selecting leader to record transactions in the blockchain network in a trusted execution environment (TEE). This protocol minimizes energy and computing power consumption. \textbf{Here, TEEs are prone to security attacks which will compromise the entire network. (QH: Please check the correctness and accuracy of this statement!)} 
%Meanwhile, Yang \textit{et al.} \cite{12} modified the Delegated Proof of Stake (DPoS) protocol to increase efficiency and remove any malicious nodes appearing in the network by introducing a downgrade mechanism. 
 % that are possible in the given consensus protocols and the persistence of the blockchain networks against those attacks.

To address the aforementioned shortcomings, we propose an efficient, permissioned blockchain system with transaction prioritization mechanism for smart cities. In general, smart cities have a very high network traffic, and there are  chances that some  critical information, such as information about an emergency vehicle's location heavily affecting the patient's health, will experience some delay before being recorded in the network, unless a transaction prioritizing mechanism is implemented to prioritize transactions based on their urgency levels instead of transaction fees or arrival time.
The robustness and persistence of our proposed permissioned blockchain network against various common security attacks are also investigated in detail.

\section{System Model}

In a blockchain-based smart city, different applications are integrated using a blockchain network. In such a scenario, many transactions will enter the blockchain network at the same time, such as information regarding street lighting and overflow of wastes, optimizing traffic lights, notification about an accident, data about $CO_2$ emissions, information of vehicles including emergency vehicles. Hence, transaction prioritization, i.e., prioritizing transactions based on their urgency level, is highly significant for the blockchain network to function smoothly and efficiently.  Without transaction prioritization mechanism to handle this traffic, the traditional blockchain network will validate transactions based on transaction fees or arrival time depending on the employed incentive mechanism, which can cause potential damages in some cases. 

In particular, there are chances that a high priority transaction, such as emergency report, might need to wait for a long time if its fee is lower or it arrives right after generating a block. For a smart city-based application in which the blockchain network serves as an interoperable platform to store data for communication, an ambulance carrying an urgent patient will be sending information regarding its location to smart traffic lights for easy passage on its way. This information from the emergency vehicle will enter the blockchain system with other transactions. A transaction with higher fee might get recorded first, if the blockchain system involves the transaction fees as incentives for blockchain nodes. This will cause the transaction related to the ambulance to wait in a mempool causing delay in optimizing traffic lights for the ambulance. In reality, for a person in urgent need of medical care, every second counts. Even a small amount of delay might affect the patient's health, leading to more serious problems. Hence, ensuring that critical or high priority transactions are not delayed or worst case lost among high fee transactions becomes significant.

\subsection{Transaction Prioritization}In this paper, by introducing transaction prioritization in the blockchain network, we aim to  prioritize transactions based on their urgency level so as to guarantee that  transactions with high urgency level are immediately dealt with and requisite actions are immediately taken by a relevant party without any  delay. Prioritizing transactions can also ensure efficiency and smooth functioning of the blockchain network. We have assumed that the government gives an insight into what transactions comes under the high priority transactions and  low priority transactions groups. Based on this, the clients will tag the transactions before broadcasting those transactions to the blockchain network. Here, we assume the clients will be honest or will undergo penalty somehow.

 Transactions with information requiring timely response, such as optimizing traffic lights for emergency vehicles, notification regarding accidents, water leak in pipes, and overflow of wastage, can be considered as priority transactions. Transactions containing general or regular information, such as updates from various sensors, notifications about road conditions, and managing street lights, can be considered as normal transactions. If there is a priority transaction in the pool, then that transaction is immediately recorded without delay, while a normal transaction in the blockchain network, generally, undergoes a  wait time. On this basis, for the example discussed above, transactions related to optimizing traffic lights will be recorded before any other transactions in the network, thus creating smooth passage for the emergency vehicle in extreme traffic conditions, thus reducing the emergency vehicle's travel time as much as possible. This will also help in reducing emergency vehicle accidents and risks involving patient's life.

Mostly, in the existing blockchain networks, blockchain nodes are usually motivated by monetary benefits they earn for mining and recording transactions; hence, they record transactions with high fees first, such as miners in Bitcoin. However, transaction prioritization mechanism focuses on general welfare of the public, by prioritizing transactions based on its urgency level without giving precedence to fees. Hence, deploying transaction prioritization in the existing blockchain networks or having an extension to the existing blockchains becomes burdensome as  interest-driven blockchain nodes  might not honestly execute transaction prioritization but focus on  only mining transactions with high fees.
\subsection{Blockchain-based Smart City Design}
To solve the above  challenge, we focus on deploying a consortium blockchain network for smart city applications with a newly designed consensus protocol to achieve the aforementioned feature of transaction prioritization. Since the network is permissioned, only nodes approved and permissioned by the management entity will participate in validation and recording transactions in the blockchain network. %Nodes participating in this blockchain network are called blockchain nodes which help in recording transactions, and  
Our system contains an efficient leader node  and multiple follower nodes. The leader node  is updated for every $b$ blocks where $b$ indicates the maximum number of blocks any leader can generate. The leader node records transactions onto the blocks based on their prioirity, while the follower nodes will verify the leader's work.
Moreover, since the nodes' rewards/penalties are related to the nodes' working quality, we can be ensured of the system working in an efficient and honest manner. Hence, the incentivizing mechanism in  our system design model ensures  honest behaviour of nodes in the network. In our system design, we also assume that all nodes in the network are synchronized which will make it easy for the followers to verify the leader's work in real-time.

Citizens can access all information recorded on this permissioned blockchain through a blockchain client interface, %. Therefore, citizens  will have complete knowledge about everything happening in the network, 
thus ensuring transparency and availability of truthful information for their reference.  With the presence of fewer nodes for validating transactions than the public blockchain, the scalability of our employed permissioned blockchain will be improved, thus being applicable to solve day-to-day problems in a smart city much more efficient and faster. The blockchain network, in general,  adds a security layer to critical records that are particularly exposed to high corruption risks, thereby strengthening the integrity of the smart city management.  
For illustration, Figure 1 describes the system model of our proposed novel consortium blockchain network for smart city applications, which consists of two parts. The right-side is the  smart city part including some of the smart city applications supported by the blockchain network. And the left box refers to the consortium blockchain network, including a leader and multiple followers for a proper and efficient governance. These nodes which are participating in the consortium blockchain network are pre-approved by the authority entities of smart cities to avoid malicious actors who may hinder the city management from joining the network.
 \begin{figure}[hbt]
\centering
\includegraphics[width=0.45\textwidth]{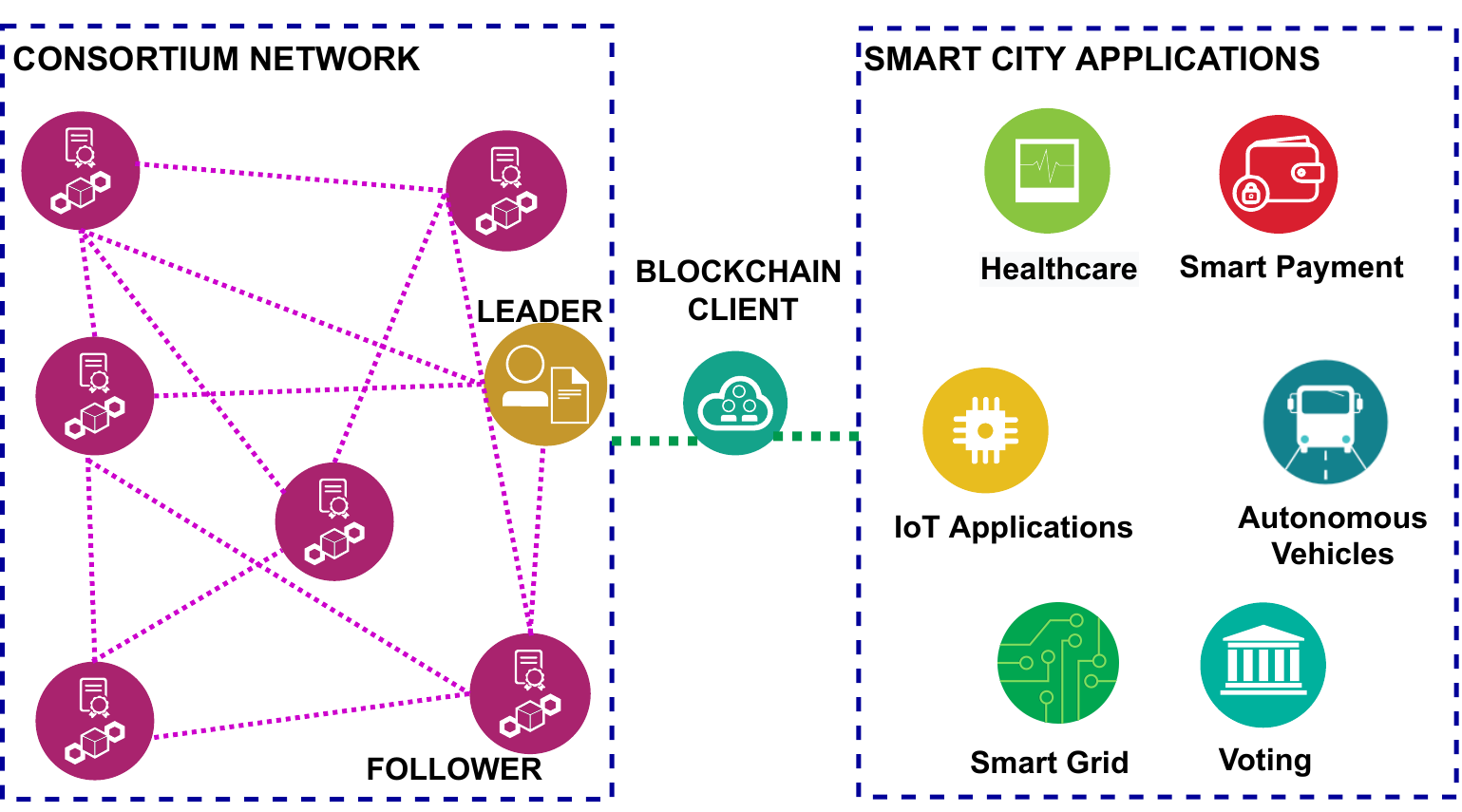}
\caption{ System model of our blockchain-based smart city application.}
\label{fig:1}
\end{figure}

\section{Consensus Mechanism Design for Transaction-Prioritized Blockchain}
To enable the implementation of our proposed transaction prioritization feature for a consortium blockchain for smart cities, we design a new consensus mechanism in this section. The overall process is introduced in Section \ref{subsec:main}, followed by detailed designs of some important steps in Sections \ref{subsec:leader} to \ref{subsec:peer prediction}.

\subsection{Overall Procedures of our Consensus Algorithm}\label{subsec:main}
\begin{figure}[hbt]
\centering
\includegraphics[width=0.45\textwidth]{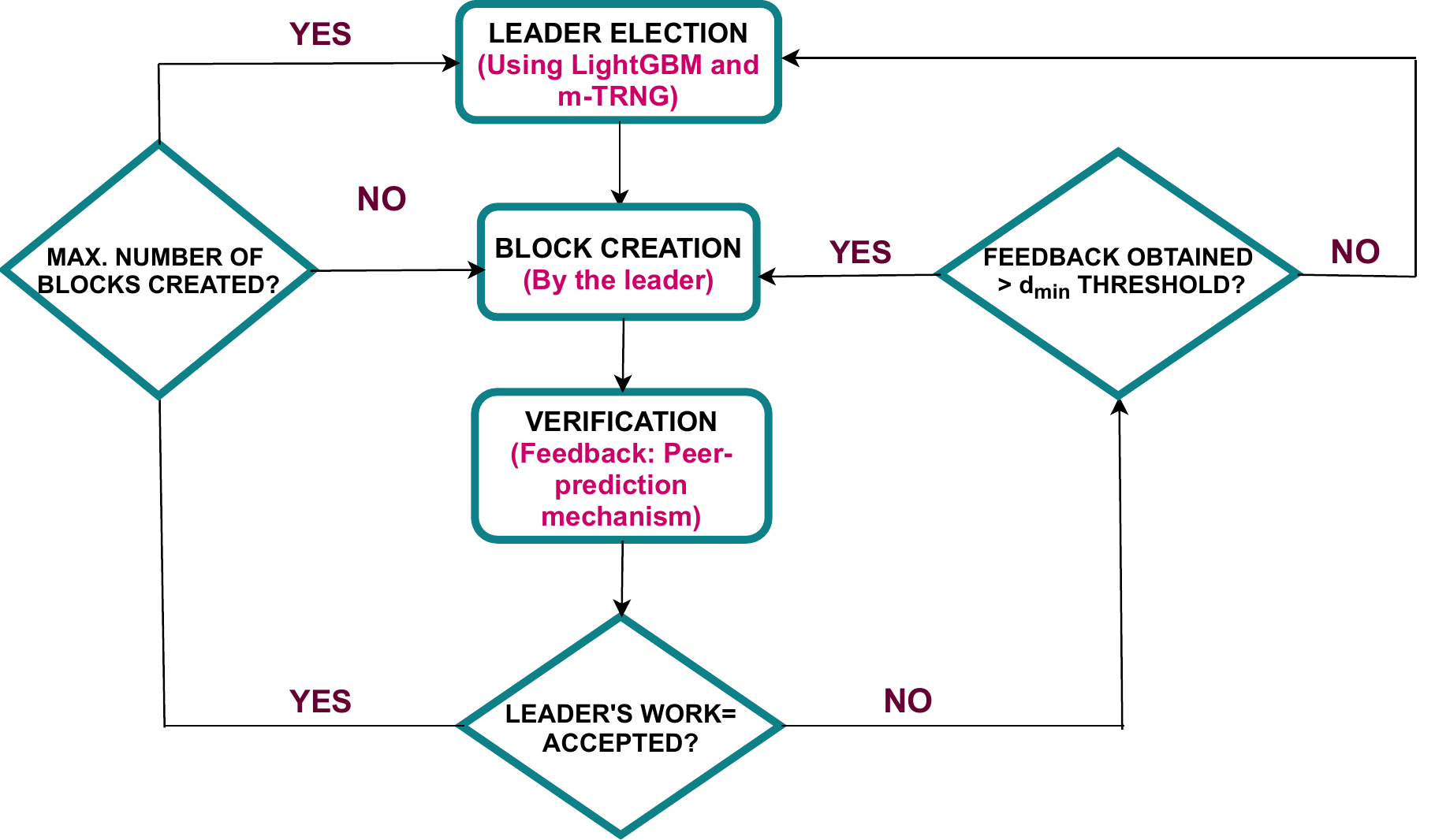}
\caption{ Flowchart of our consensus protocol for blockchain-based smart city applications.}
\label{fig:2}
\end{figure}

Our consensus mechanism consists of four significant steps to properly record transactions that are entering the blockchain network. The overall process is illustrated in Figure \ref{fig:2} and can be explained as follows:
\begin{enumerate}
    \item \textbf{Leader election.} Leader election becomes a critical part in our consensus mechanism since selecting a trustworthy and efficient leader ensures that the process of prioritizing transactions is carried out in an honest and efficient manner without the leader involving in any kind of malicious activities. Initially, a leader is chosen among blockchain nodes for a specific time period to generate  new blocks in the network. During that time, other blockchain nodes will act as followers. To choose the leader in a fair manner and also to ensure that an efficient node is selected as the leader, LightGBM, an machine learning (ML) algorithm, and a modified true random number generator (m-TRNG) are employed, which will be elaborated in Section \ref{subsec:leader}.
    \item \textbf{Block creation.} In our consensus algorithm, the leader is required to immediately record all the priority transactions that are entering the network at that specific time. Normal transactions that are currently present in the pool along with the priority transactions are recorded without delay. Other normal transactions will experience some delay before they are recorded. To this aim, we design a dynamic block creation algorithm in Section \ref{subsec:block}.
    \item \textbf{Verification.} The newly generated block  is then broadcast to the whole blockchain network to get verified. Followers check the work of the leader after every block creation for examining whether the leader follows the transaction prioritization rule. Feedback from followers are collected using a peer prediction mechanism \cite{1} for truthfulness consideration. Followers are motivated to give honest feedback to obtain high trustworthiness scores, which will subsequently bring high rewards for them during incentive distribution step. 
    %\item \textbf{Decision on the leader's work.} 
    Based on the followers' review, the block is either determined to be accepted and then appended to the main chain or cast aside. A new round of leader election will also be triggered under specific conditions. In detail, if the review for a new block is satisfactory, then the leader continues to record the next block till the maximal number of blocks has been reached. If the review score is unsatisfactory and below the minimum threshold ($d_{min}$),  the leader is voted out and this will put forth the next leader election process in motion. All nodes in the blockchain network will be notified about the election and the reason behind it via gossip protocol \cite{15}. Also, if the work is unsatisfactory but between the range of minimum ($d_{min}$)  and maximum ($d_{max}$) thresholds, then the work is rejected but the leader is given another chance to produce a new block for error tolerance. 
    The detailed design of this part is presented in Section \ref{subsec:peer prediction}.
    \item \textbf{Incentive distribution.} Incentive mechanism for our system is based on the trustworthiness scores for followers and the feedback on work quality for the leader. Incentives are distributed in the blockchain network when the managing entity of this network inputs incentives into the blockchain system. Since the trustworthiness of the followers and the number of generated blocks of the leaders will be recorded on the main chain, there is enough evidence to achieve fairness in incentive distribution.  
\end{enumerate}

% Initially, a leader is chosen among blockchain nodes for a specific time period to generate  new blocks in the network. During that time, other blockchain nodes will act as followers. To choose the leader in a fair manner and also to ensure that an efficient node is selected as the leader, LightGBM, a machine learning (ML) algorithm and a modified true random number generator (m-TRNG) are employed. 

% Figure 2 indicates the flowchart of our novel blockchain network.

\subsection{Leader Election}\label{subsec:leader}
The first critical step in our novel consensus protocol for the consortium blockchain-based smart city is the  leader election process, which comprises of two stages: determining efficient leader candidates by obtaining a candidate list, and then, randomly selecting a leader from that list. First, the candidate list is obtained using LightGBM algorithm \cite{2}, where $n$ candidate leaders are chosen. These candidates are selected based on the parameters which are closely related to the blockchain nodes'  efficiency and trustworthiness. Here, the number of candidates chosen is far less than the number of blockchain nodes. Given this candidate list, the final leader and the maximum number of blocks the leader can create during its session i.e.,  $b$  $\epsilon$ $[1, b_{max}]$   will be obtained using a random function \cite{3}. This random number $b$ is known only to the leader and other candidates in the list but not to other nodes not in the candidate list for security purposes. 

\subsubsection{LightGBM}
In our consensus protocol, every trustworthy node in the network will have a chance to become a leader to maintain fairness to avoid distributed denial-of-service (DDoS) attacks. When a leader is selected, other nodes act as followers for that specific round. Hence,  we deploy LightGBM, an ML algorithm which is based on a gradient boosting framework and is more scalable and accurate than other gradient boosting trees is used. 
The algorithm also implements a highly optimized histogram-based learning algorithm, resulting in higher efficiency and
less memory usage.

Before deploying our ML algorithm, four critical parameters are considered to determine efficient leader candidate nodes. These parameters  comprise of: 
\begin{enumerate}
\item Trustworthiness score (denoted by $T$),
\item Number of  peers ($P$),
\item Efficiency value ($E$), and
\item Vote-outs ($V$) if any.
\end{enumerate}   
To be specific, $T$ indicates the honesty of every follower when submitting feedback on the leader's work; $P$ refers to the degree of the node i.e., the connectivity of the node in the blockchain network; $E$ is the efficiency of a node as a leader and is given by 
\begin{equation}
 E= BCT - \tau,  \nonumber 
\end{equation}
where $BCT$ represents the block creation time and $\tau$ indicates the time at which the last transaction in this block was recorded; And the history of $V$ reflects how inadequate of this node being a leader in previous experiences.
%is a setback for the node as it means that the performance of the node as a leader is not adequate.

Among these four parameters, the first two parameters are retrieved when a blockchain node acts as a follower, while the remaining parameters are related to the leader role of the node. Information about a node's trustworthiness score and number of peers are obtained via a smart contract triggered after the peer-prediction mechanism. Variables $BCT$, $\tau$, and hence the value of $E$ can be obtained from the blockchain (using timestamps). If the leader does not create the next block on time and if there is a time lag in recording transactions, the value of variable $E$ will suffer. If a node acting as a leader is voted out,  that information will be broadcast to all consortium members, so this information is visible to all nodes present in the network. These parameters are usually updated after each block is either accepted or rejected based on the followers' review. Table 1 gives an illustration of different node parameters and whether they are associated with a leader or a follower role.

\begin{table}[htb]
\caption{Node parameters}
\centering
\begin{tabular}{|c||p{1cm}<{\centering}|p{1cm}<{\centering}|p{1cm}<{\centering}|p{1cm}<{\centering}|} 
\hline
Attribute & $T$ & $P$ & $E$ & $V$\\%[1.0ex]
%\hhline{|=====|}
\hline
Tag & \multicolumn{2}{c|}{Follower} &
\multicolumn{2}{c|}{Leader} \\
\hline
\end{tabular}
\label{tab:node_parameter}
\end{table}

The afore-mentioned parameters are obtained instantly from the blockchain network, and by involving these parameters, LightGBM can output a leader candidate list comprising of $n$ candidate nodes.  Since these parameters are incorporated in determining candidate nodes, the candidate nodes will be more trustworthy and efficient than other nodes in the network. Once this list is obtained, randomness is incorporated as the second step of the leader election process using m-TRNG function \cite{3}, to select the next leader.  %The maximum threshold for this random number is $b$ blocks. 

In our leader election process, this LightGBM algorithm is run by the current leader  of the blockchain network  to determine the candidate list for next round of leader election.
In case the current leader is voted-out before running this ML algorithm for the next round of leader election, a candidate node with highest trustworthiness score value will run this algorithm instead to generate the next candidate list. Voting-out occurs only if the followers' final decision ($D$) was to reject the leader's work. We introduce thresholds like $d_{min}$ and $d_{max}$ to make decision on leader's work based on followers' opinions. If $D \leq d_{min}$, then the leader is voted-out ($V=1$). Here, $d_{min}$ represents a minimum threshold value below which the leader's work is rejected.  For other values of $D$, the leader continues to generate till $b$ blocks.

\subsubsection{m-TRNG}
After the candidate list is determined, the current  leader or a candidate node with a high trustworthiness score (in-case if the leader is voted-out) utilizes a modified true random number generator (m-TRNG) based on the entropy of the network to complete the next leader election process by generating true random numbers that cannot be predicted easily. This entropy will depend on the unconfirmed transaction pool in the network. Due to the dynamic nature of this pool's size, it can be used to generate a true random number. Since our m-TRNG is based on the entropy of the blockchain network, the increase in incoming network traffic will raise the level of randomness, thereby causing the output to be less predictable. %This will ensure security in the blockchain network, making m-TRNG a vital component in our blockchain network.

In our system, m-TRNG function will start  as soon as the candidate list is determined using ML, and it will output two true random numbers to indicate the leader selected and the maximum number of blocks that the leader can generate ($b$). The above-mentioned feature of entropy-based m-TRNG enhances the security of the blockchain network with respect to resisting DDoS attacks. Detailed steps involved in the leader election process are shown in Algorithm 1.

We first initialize both the  current leader's vote-out value ($V_{C}$) and final decision ($D$) of the followers (Line 1).  Before the current leader generates the maximum number of blocks  it is supposed to generate ($b_{C}$), i.e., when its generated block number $(b_{CB})$ is within $b_{C}$ value, the next leader needs to be determined (Line 2). In case, before the current leader runs LightGBM, an ML algorithm  to determine the next leader, if the current leader's work is rejected by the followers, i.e., if the final decision of the followers ($D$) becomes less or equal to a minimum threshold below which the leader's work will be rejected ($d_{min}$) then the current leader is voted-out ($V_{C}=1$). Then the candidate node with high trustworthiness score value ($T$) will run the ML algorithm to 5determine the candidate nodes (Lines 3-6).  Otherwise, the current leader can continue generating blocks since its work has been accepted by the majority of the followers. Hence, the current leader's vote-out value will be $0$ and the leader can run LightGBM to determine the next leader candidate nodes (Lines 6-9).
This LightGBM algorithm will calculate trustworthiness score ($T$), number of  peers ($P$), efficiency value ($E$), and vote-out value ($V$). Based on these critical parameters, $n$ candidate nodes are determined. Among these $n$ leader candidate nodes, the next leader node is selected by a random function generator, namely, m-TRNG. This random function also specifies the maximum number of blocks the next leader can generate ($b_{N}$) (Lines 10-13).

\begin{algorithm}[h]
\caption{Leader Election}
\label{alg:algo1}
\begin{algorithmic}[1]
\Require Trustworthiness Score ($T$), The number of Peers ($P$), Efficiency Value ($E$),  Vote-outs ($V$), The current leader's vote-out value ($V_{C}$), The current leader's generated block number ($b_{CB}$), The maximum number of blocks the current leader can generate ($b_{C}$), Decision of the followers on the current leader's work ($D$), Minimum threshold value of $D$ below which the leader is voted-out ($d_{min}$)
\Ensure A new leader
\State Initialize $V_{C}=0, D=0$
\While {$b_{CB} \leq b_{C}$ } 
\If{$D \leq d_{min}$}
 \State  $V_{C}=1$ and the current leader is voted-out
    \State The current candidate node with high $T$ value runs LightGBM followed by m-TRNG
 \Else
 \State  $V_{C}=0$
 \State The current leader runs LightGBM followed by m-TRNG
    
    \EndIf
 \State This LightGBM algorithm obtains the values of $T$, $P$, $E$, and $V$ for all nodes from the network
 \State $n$ candidate nodes are selected by the algorithm
\State  m-TRNG runs on these selected $n$ candidate nodes
\State Next leader and the number of blocks to be generated by that leader is randomly selected
    % \Else
        %\If{$T_{C}$ $\geq$  $w$}
        % \State Block is created with timestamp 
        %\Else
        % \State Break
    % \EndIf
\EndWhile
\end{algorithmic}
\end{algorithm}
\subsection{Block Creation}\label{subsec:block}
In the second step, the leader records transactions presenting in the mempool onto the main chain by following certain rules. That is, while creating a block, priority transactions are given more significance. Hence, when there is even one priority transaction appearing, the leader needs to immediately create a block and records this transaction along with other normal transactions in the mempool, which is then broadcast to the blockchain network. This rule will ensure that priority transactions do not encounter any delay.  Sometimes, however, when there are limited number of transactions entering the network, there might be some unused spaces in the blocks created. On the other side, normal transactions might undergo some delay in this process. % as more attention is given to the priority transactions.

To achieve this goal, we design a dynamic block creation policy for the leader to properly create blocks and record transactions. 
Detailed steps involved in the block creation process are shown in Algorithm 2.
\begin{algorithm}[h]
\caption{Dynamic Block Creation}
\label{alg:algo1}
\begin{algorithmic}[1]
\Require Maximum number of transactions can be recorded in a block ($m$), number of normal transactions ($n_{t}$), number of priority transactions ($p$), current waiting time ($T_{C}$), and the maximum waiting time ($w$)
\Ensure A new block
\State Initialize $T_{c}=0$
\While {true} 
\If {$p$ $\geq$ $1$}
 \State Block is created with timestamp
 \Else
    % \If{$T_{C} <  w$}
        \If{$n_{t}$ $\geq$ $m$ or $T_{C}$ $\geq$  $w$}
            \State Block is created with timestamp 
        \EndIf
    % \Else
        %\If{$T_{C}$ $\geq$  $w$}
        % \State Block is created with timestamp 
        %\Else
        % \State Break
    % \EndIf
\EndIf
\State Wait for followers' feedback
\EndWhile
\end{algorithmic}
\end{algorithm}

At any moment, the presence of even one priority transaction in the pool will lead to the creation of block immediately (lines 3-4). Here, all priority transactions presented in the mempool will be recorded in the block followed by normal transactions. If there is no space to record all normal transactions, then the leader will leave some waiting transactions for the next block creation. 
If all transactions in the mempool are normal transactions, then the transactions are recorded based on their arrival time, i.e., a normal transaction $n_{t1}$ is recorded before a normal transaction $n_{t2}$ if $n_{t1}$ has entered the network first. Block creation time for normal transactions depends on whether there are enough normal transactions in the mempool to entirely occupy a block. If the current number of normal transactions in the pool is equal to or larger than $m$, or if the time elapsed is equal to or larger than $w$, then  a block is created (lines 6-7). % if $t$ is  equal to or greater than $m$; be. 
This will ensure a tradeoff between the block space utilization and system efficiency on packaging priority transactions. %that no space in the block is wasted, and avoids empty blocks being created by the leader. 
%Also, to ensure that there is no unacceptable waiting period for normal transactions, there is a wait time (say $w$) (line 9-10) after which a block is created even though $t$\textless$m$. 
After finishing creating a block, this block is broadcast to the network to let the followers review the work of the leader. The leader needs to wait for the feedback of that block from the followers (line 10) in the network before it can create a new block.

Thus, the block creation time in our consensus mechanism is dynamic which depends on whether there exists any priority transaction or enough normal transactions. If a transaction is a priority transaction, a block including it is created instantly. However, in case of normal transactions, a block  is created only if there are enough normal transactions to occupy the block till the maximum waiting time $w$. After $w$, a block is created even if there are not enough normal transactions to avoid endless waiting of transactions. 

%After creating each block, that block is broadcast to the network to let the followers review the work of the leader. If the work of this block is fair after evaluated by every follower, then the block is appended on the main chain, else the leader is voted out and the block is cast aside. The leader waits for the followers' review before creating the next block. 

\subsection{Peer Prediction-based Feedback Collection}\label{subsec:peer prediction}
After creating a block, the leader broadcasts the block in the network for verification purposes. Followers need to review the leader's work and provide honest feedback. Since some follower nodes might become malicious to thwart the governance of the smart city,  a peer-prediction mechanism is employed to guarantee the  trustworthiness of the feedback from  followers to ensure true feedback from the followers.  %Here, both priori and posteriori predictions of the followers are needed to ensure that the followers are behaving honestly. 
As truthful feedback from followers are very critical,  incentive rewards based on trustworthiness scores will be provided to motivate honest followers. To protect the privacy of followers, the feedbacks provided by  followers are not directly related to the leader's work, but are inferred from the collected predictions for other peers. For example, follower $i$  predicts its peer $j$'s opinion, from which $i$'s feedback can be inferred.

In our scenario, we denote the number of followers in the blockchain network as $N$ and the quality of a leader's work $W$ is evaluated as either $a$ or $r$, where $a$ represents acceptance and $r$ represents rejection for the leader's work from the feedback of any follower. Each follower $i \in \{1,2,\cdots,N\}$ reviews the leader's work, and then makes an opinion of the review which is denoted by $S_{i} = s_{i} \in \{a, r\}$.  

Two key parameters in  followers' posterior beliefs are \textit{false alarm probability} and \textit{missed detection probability}. The false alarm of the judgement is noted when a work 
is misjudged and falsely rejected, when actually it is an acceptable work (similar to false negative). A follower $i$'s false alarm probability of judgement is denoted by $P_{fa,i} = P (S_{i} = r | W = a )$. On the contrary, a missed detection
probability of judgement is noted in case when a follower accepts the  work when in fact, the work is poor (similar to false positive). Missed detection probability of judgement is given as $P_{md,i} = P (S_{i} = a | Q = r )$. 
 
To obtain prior beliefs and posterior beliefs of the followers \cite{1}, initially, every follower $i$ is paired with another follower as its peer (denoted as $j$), and is required to report the peer's prior and posterior beliefs of the leader's work before and after the block is broadcast. Hence, the prior belief for a block $B$ is obtained  when the leader is recording transactions in $B$. After the leader broadcasts $B$, the followers are requested to submit their posterior beliefs. Based on these two reports, the follower $i$’s trustworthiness can be calculated by a proper scoring rule which indicates if $i$ is honest or not.
\subsubsection{Prior Belief}
Each follower $i$ has one peer $j$ selected randomly from other followers,  and $j$ is required to review the same block $B$. With no knowledge about the leader's work quality, follower $i$ is required to report its prior belief $y_{ij} = [0, 1]$, denoting the probability that its peer $j$ will accept the leader's work, i.e., $x_{j} = 1$. Henceforth, $y_{ij}$ is given by,
\begin{equation}
    y_{ij} = P_{i} ( x_{j} = 1),
\end{equation}
which can be calculated in detail as,
\begin{equation}
  \begin{split}
     y_{ij} = P_{i} (x_{j} = 1 |W = a)P_{i} (W = a)+\\ P_{i} (x_{j} = 1 |W = r)P_{i} (W = r).
  \end{split}  
\end{equation}
Here, $P_{i}$($x_{j} = 1 |W = a)$ represents the probability that  follower $j$ gives a report of accepting the leader's work, when user $i$ makes acceptance judgement to the same work. This judgement is private, and only $i$ knows this information.
\subsubsection{Posterior Belief}
In this report, follower $i$ should make its opinion and send posterior belief to the network after reviewing the actual work of the leader. Here the posterior belief is denoted by $y_{ij}' (s_{i}) = [0, 1]$, indicating the probability that $i$'s peer follower $j$ will accept the work. Then $y_{ij}'$ is given by
\begin{equation}
    y_{ij}' (s_{i}) = P_{i} (x_{j}=1 | S_{i}=s_{i}),
\end{equation}
and  can be decomposed into two conditions as follows
\begin{equation}
    y_{ij} (r) =\frac{a_{1}(1 -P_{f,j}) + a_{2} P_{m,j}}{a_{1}+a_{2}},
\end{equation}
\begin{equation}
     y_{ij} (a) =\frac{a_{3}(1 -P_{f,j}) + a_{4} P_{m,j}}{a_{3}+a_{4}},
\end{equation}
where $a_{1}=P_{fa,i} P(W = a)$, $a_{2}=(1-P_{md,i})P(W = r)$, $a_{3}=(1-P_{fa,i}) P(W = a)$, and $a_{4}=P_{md,i} P(W = r)$. 

 In prior belief, even before $B$ is created, follower $i$ predicts $x_{j}=1$ with a probability. Later, if the leader's work is truly satisfactory, the probability of $i$ predicting  $x_{j}=1$, after reviewing the work, will be  higher than its prior belief. Since, in posterior belief, $i$ has a knowledge about the quality of leader's work which was absent in prior belief. Hence, if the work is satisfactory, follower $i$'s posterior belief  will be greater than its prior belief. Similarly, $i$'s posterior belief will be lesser than its prior belief if the leader's work is not satisfactory. 
\subsubsection{Scoring Rule and Trustworthiness Value}
Posterior belief is followed by a proper scoring rule for calculating a trustworthiness score for each follower, which motivates followers to provide honest feedback. Scoring values are assigned to the  followers based on their honest feedback. Here proper scoring rules, such as logarithmic, quadratic, and spherical scoring rules, can encourage followers to  provide their feedback truthfully to increase their rewards.  Logarithmic scoring generates negative values, which will, in our application, result in dishonest followers getting penalties. To avoid this problem and to motivate followers in a positive way, %logarithmic scoring rule is avoided in our mechanism. Both quadratic and spherical scoring values differ only by an additive constant when scaled to induce same precision. Since there is essentially no difference in the results elicited from followers, in our scenario, 
we use quadratic rule in this paper. The followers  need to provide their feedback $y \in [0,1]$ truthfully  to maximize their trustworthiness scores and accordingly their rewards. The binary quadratic scoring rule ($R_{qi}$) for follower $i$  is given by
\begin{equation} \label{eq1}
\begin{split}
R_{qi} (y, \omega = 1) & = 2y-y^{2} ,\\
 R_{qi} (y, \omega = 0) & = 1- y^{2},
\end{split}
\end{equation}
where %subscript $q$ denotes the scoring based on the quadratic rule, and 
$\omega \in \{0 ,1\}$ indicates the binary report submitted to the blockchain network. %, and $y$ ranges between $0$ to $1$.
%$R_{qi}$ denotes the score value of follower $i$ for the feedback it sent, and the score is based on the quadratic scoring value.
Trustworthiness is positively dependent on the scoring value. If a follower has a low trustworthiness value, then its opinion would not be taken into account as there are high chances that the follower is acting maliciously. The trustworthiness of a follower $i$ is therefore defined as
\begin{equation}\label{trust}
     T_{i} =\alpha R_{qi} + (1-\alpha)\hat{T_{i}} + 1-\beta_{i}.
\end{equation}
In the above equation, $\alpha$ is a weighted parameter ranging between 0 to 1, which indicates putting more the significance on the current trustworthiness score value or previous trustworthiness values $\hat{T_{i}}$. % is its previous trustworthiness score value. 
Also, $1- \beta_{i}$ denotes the promptness index of follower i, and the latency ($ \beta_{i}$) is given by
$ \beta_{i}= \frac{\text{Latency of follower i}} {\text{Latency range}}$, where the denominator is the difference between the maximum and minimum latency values observed in the network. The value of $\beta_{i}$ ranges between 0 to 1. 
It is used to determine how quickly the followers are able to complete their review work. 
\subsubsection{Inferred Opinion}
Follower $i$'s opinion on the leader's work is inferred from its prior and posterior beliefs in which  follower $i$ has predicted its peer follower $j$'s opinion. Then, the formula to infer $i$'s opinion on the leader's work is given by,
\begin{equation}
       x_{i} = x(y_{ij}, y'_{ij}) =
    \begin{cases}
 1 & { y'_{ij} > y_{ij}},\\
  0  & { y'_{ij} < y_{ij}}.
\end{cases}   
\end{equation}
If the posterior belief $y'_{ij}$ is greater than the prior belief $y_{ij}$, follower $i$ has accepted the leader's work; while if the posterior belief is less than the prior belief, then the follower is not satisfied with the leader's work. 
The final decision, denoted as $D$, of the leader's work  is given by the value of  
\begin{align}\label{eq:D}
   D=\sum_{i=1}^{h} \frac{x_{i}}{h},
\end{align}
where $x_{i}$ represents trustworthy nodes' opinions on the leader's work. Trustworthy nodes are nodes whose trustworthiness score is greater than 50 \%, and $h$ represents total number of trustworthy nodes in the network. Only these trustworthy nodes' opinions are taken into consideration about evaluating the leader's work.
Then, three possible results of a decision will be made according to where $D$ lies in: 
\begin{enumerate}
    \item If $D\leq d_{min}$, the leader's work is rejected and the leader is voted out; 
    \item If $d_{min} \leq D\leq d_{max}$, the generated block will still be rejected but the leader will be given another chance to generate a new block as a sort of fault tolerance; 
    \item If $D \geq d_{max}$, the leader's work will be directly accepted and appended to the main chain.
\end{enumerate} 
%  \scalebox{0.76}{
% \begin{align}
%  D=\sum_{i=1}^{h} \frac{x_{i}}{h}
% \begin{cases}
%  \geq min,  \text{work rejected and leader is voted-out},\\
%  \geq max, \text{work accepted}, \\
%  otherwise, \text{work rejected but leader continues}.\\
% \end{cases}   \notag
% \end{align}

In the above, $d_{min}$ and $d_{max}$ represent different thresholds used to make a decision on the leader's work. 
Based on the decision, the leader is either voted out or continues to create the next block if it is yet to attain the goal of creating the maximum number of blocks. In this peer-prediction mechanism, two smart contracts are used as an implementation method for collecting prior and posterior beliefs from the followers. The first smart contract is triggered before $B$ is created by the leader, and the second smart contract is triggered after $B$ is created to obtain the followers' true opinions.

To summarize, we present specific steps involved in the above peer prediction-based verification and decision making process in Algorithm 3.

\begin{algorithm}[h]
\caption{Peer-prediction Mechanism}
\label{alg:algo1}
\begin{algorithmic}[1]
\Require $y_{ij}$, $y_{ij}'$, $R_{qi}$, number of followers ($N$)
\Ensure Verification of leader's work
\State Initialize variable $T_{i}=0$
\While {true} 
\For{$i$ $=$ $1$ to $N$}
 \For{$j$ $=$ $1$ to $N$}
 \State $y_{ij}$ for $B$ is obtained
 \State Wait till $B$ is created and broadcast
 \State $y_{ij}'$ for $B$ is obtained
 \State $R_{qi}$ is determined based on quadratic rule
 \State $T_{i} =\alpha(R_{qi}) + (1-\alpha)T_{pi} + 1-\beta_{i}$ 
 \State $x_{i} = x(y_{ij}, y'_{ij})$
  \State Decision ($D$) on the leader's work is made
    \EndFor
 \EndFor
\EndWhile
\end{algorithmic}
\end{algorithm}

At the end of each day, the leader and followers participated in the entire consensus process are motivated by rewards  inputted by smart city governors. The rewards for followers will be positively proportional to their trustworthiness value. % $Reward = f(T_i)$. 
While the rewards for the leader will mainly be calculated according to the number of blocks appended on the main chain; also, it may include transaction fees associated with normal transactions. %The incentive mechanism is dynamic based on the inflow of money.

\section{Security Analysis}
The greatest advantage of any blockchain network is its promise of security and privacy due to its decentralized nature. Blockchain  also  improves  confidentiality and creates trustful environment for different parties to work together. %In today's world, newer security attacks are emerging each day, targeting vulnerabilities in the blockchain network. 
To investigate the security performance of our proposed consensus protocol, we analyze several typical security attacks in blockchain, such as majority attacks, DDoS attacks, replay attacks, and empty block attacks.
\subsection{Majority attack}
Although the majority attack, i.e., $51\%$ attack, is  relatively difficult to launch, such an attack will result in severe outcomes if it happens on the blockchain network. A majority attack might happen in our network as well when a blockchain node in the network colludes with other nodes and gains more than $50\%$ of control over the whole network, after which it has the power to destroy the proper functioning of the blockchain-based smart city. It will also render our blockchain network to be inefficient. However, we design our proposed blockchain network  in such a way to prevent such an attack from the following three aspects:
\begin{enumerate}
    \item Our blockchain network is permissioned, and thereby, only certain nodes which are approved by a management entity supporting general welfare is allowed to participate in the network.
    \item Also, the reports of followers are collected based on the peer-prediction technique, making it not beneficial for nodes to collude with each other. 
    \item Due to the truthful results collected via the peer prediction-based mechanism, if a supposedly trustworthy and efficient blockchain node tries to slack in its duty or act in a malicious manner after being selected as the leader, that blockchain node will be voted out immediately, which will be recorded on the blockchain permanently, leading to reduced incentives and decreased probability of being selected as a leader in the future.
    
\end{enumerate}

\subsection{DDoS attack}
DDoS attack is one of the common attacks in the blockchain network, which targets the victim server to make an attack on the network by bringing huge network traffic from multiple distributed resources, and consequently, shutting down the victim server. In our blockchain-based smart city application, such an attack is possible, even though the blockchain network is decentralized, the leader node can be an obvious target of DDoS attack. %It gives malicious actors chances to launch such types of DDoS attack.

Our blockchain network rectifies this situation by introducing randomness in the leader selection process. In detail, the leader is selected randomly from the candidates list consisting of $n$ nodes, and so any pre-planned DDoS attack on the leader (yet to be elected) is impossible to make. 

Besides, there seems another aspect of vulnerability for DDoS attack in our system due to the design of multiple number of blocks allowed to be generated by the leader for efficiency consideration. However, we also randomly generate the maximum number of blocks a leader can record, and this number is known only to the other trustworthy candidates in the list. Hence, any malicious actor trying to attack the leader will not have any prior knowledge about the maximum number of blocks the leader can record. It means that any block the leader is creating, at a specific time, could be the final block of the leader, and attacking the current leader without knowing the maximum number of blocks it can create could be futile. If a candidate node who has the prior knowledge about $b$ turns to be a malicious attacker, DDoS attack is still not guaranteed as the malicious node may not able to predict if the feedback on the leader's work for that block is going to be accepted or rejected.

\subsection{Replay attack}
Replay attacks are  considered to be one of the sophisticated attacks to target blockchain network. Sometimes, a blockchain network will go through protocol changes or upgrades known as hard forks due to this attack, which will cause a split in the blockchain network, with one side working with the knowledge of the previous protocol and other side working on a new protocol. Thus, malicious nodes may further exploit this vulnerability to achieve higher revenues. 

In our blockchain-based smart city scenario, this issue is avoided as we employ the permissioned blockchain and all nodes in the network can join only with the certification of a single management entity who focuses on the betterment of the society. Hence, when there is a protocol change, it will be reflected by all nodes supporting that entity. If there are any invalid transactions or mishaps, other followers or citizens who have access to data in the network can rectify the situation by mentioning the situation.

\subsection{Empty block attack}
Empty block generation is a common operation in a blockchain network and an attack based on that general operation may cause disruption in the blockchain network making the blockchain network almost unusable. Our blockchain network is protected against this attack by the dynamic block creation policy, which ensures that the leader cannot create empty blocks for its own interest. If a leader focuses only on its own interests, and tries to create empty blocks for the sole purpose of rewards, then the block will also be not accepted and the leader can be voted out immediately. The leader cannot even obtain rewards for the empty blocks it has created, as this rejected block will not be recorded on the main chain so that this leader will not receive the corresponding rewards. %all the blockchain nodes participating in the process will be rewarded only after $x$ blocks are recorded in the main chain. Moreover, its trustworthiness score value will also be affected very badly.

 In general, a leader cannot act in a malicious way, since as soon as the leader starts acting maliciously, the leader can be voted-out immediately by the follower nodes and will lose out its rewards. Similarly, the follower nodes cannot act maliciously, since the follower nodes' trustworthiness scores will be severely impacted which will also affect their rewards and their opinions will not be taken into account. Thus, from the above analysis, we can conclude that our blockchain-based smart city provides security against most of the common attacks on the blockchain network.

\section{Experimental Evaluation}\label{sec:experiment}
In this section, we evaluate our proposed consensus protocol, mainly analyzing the performance of the employed LightGBM algorithm for leader election and the peer prediction-based feedback collection mechanism. % and study the impact of different node parameters on the prediction of leader candidates. Also, the impact of different parameters on the trustworthiness value of the followers in peer-prediction mechanism is observed. Further, the trustworthiness values of both trustworthy nodes and malicious nodes is analyzed.

\subsection{Leader Election}
\subsubsection{Performance Analysis}
Simulated data samples from 1000 nodes are obtained and loaded as an input dataset to the LightGBM algorithm. Dataset is divided into training data and test data,  with a test size selected as 0.6. Hence, 600 data samples are treated as test data. Data is then pre-processed to clean and organize.

Parameters should be tuned to obtain good accuracy. Here, the application is mentioned as `Classification' to classify the blockchain nodes as candidate nodes to become the next leader or followers. The metric parameter is chosen as logloss since this function can determine how good our model is in predictions. The other parameters and their values are shown in Table II.  The accuracy obtained by our algorithm is 99.3\%. Accuracy is calculated using the formula, 
\begin{equation}
    \text{Accuracy} = \frac{\text{Number of correct predictions}}{\text{Total number of predictions}}.
\end{equation} 

\begin{figure}[hbt]
\centering
\includegraphics[width=0.4\textwidth]{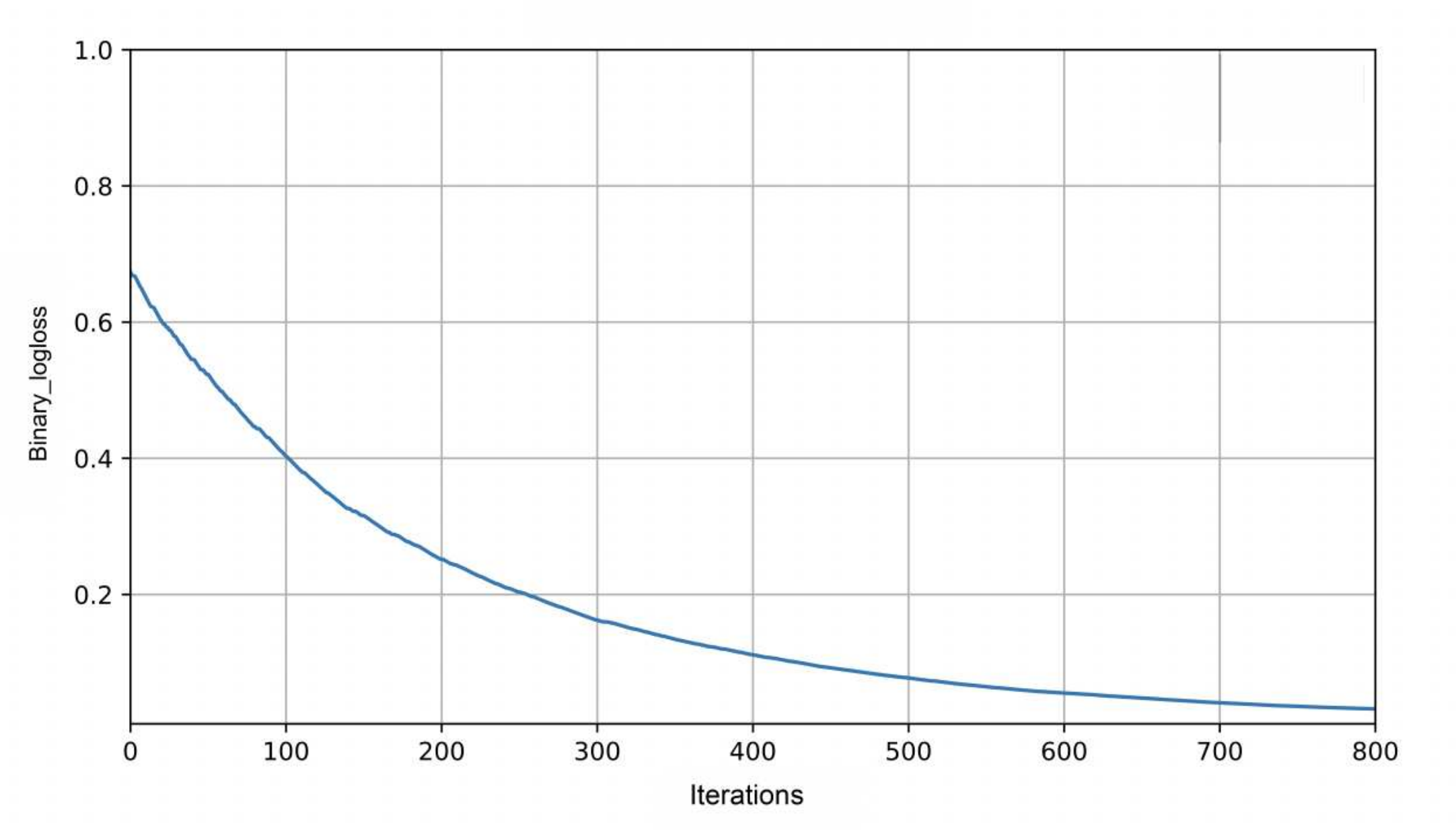}
\caption{Evolution of binary\_logloss function.}
\label{fig:4}
\end{figure}

Figure \ref{fig:4} represents the values of binary log-loss function, and it can be seen that the values converge around 0.03, thus making the algorithm more accurate. Since the metric is a loss function, the lesser the value, the greater the accuracy. 

\begin{table}
\caption{Parameters Tuning}
\centering
\begin{tabular}{|c||c|c|c|c|} 
\hline
Parameters & Values\\
\hline
Boosting type & Gradient Boosting Decision Tree \\
\hline
Learning rate & 0.005 \\
\hline
Objective & Binary \\
\hline
Metric & Binary\_logloss \\
\hline
max\_depth & 100\\
\hline
\end{tabular}
\label{tab:social1}
\end{table}

\begin{table}
\caption{Confusion matrix}
\centering
\begin{tabular}{|c||c|c|c|c|} 
\hline
Possible Outcomes & Values\\
\hline
True Positive & 356\\
\hline
True Negative & 240 \\
\hline
False Positive & 2 \\
\hline
False Negative & 2 \\
\hline
\end{tabular}
\label{tab:social}
\end{table}

\subsubsection{Impact of different parameters}
In this ML algorithm, four node parameters play an important role in predicting leader candidates. They are the number of peers, the trustworthiness score, the number of blocks generated, and the history of vote-outs. Among these parameters, the history of vote-outs has more weights. If there is a history of that particular node  been voted-out, then there are chances that that node has acted in a malicious manner or was slacking in its duty. Hence, to ensure the leader elected is always efficient and trustworthy, any nodes with history of vote-outs are given very less probability in being selected as the leader again.

The parameter with next higher weightage is the trustworthiness score, as the leader in our consensus algorithm has an important role to play in recording transactions. Also, if a blockchain node has a higher trustworthiness score, then the node has a higher probability in becoming the leader, and subsequently, their number of blocks generated will also increase. Based on the afore-mentioned four parameter values, a threshold value is computed. If the blockchain nodes have values greater than this threshold, then the nodes are considered to be  leader candidates and vice-versa.

To demonstrate the impact of these parameters on the decision of leader candidates list, we have taken a node with following parameter values as shown in Table IV.
\begin{table}
\caption{Data sample}
\centering
\begin{tabular}{|c||c|c|c|c|} 
\hline
Parameters & Values\\
\hline
Number of peers & 800\\
\hline
Number of blocks generated & 5 \\
\hline
Trustworthiness score & 1 \\
\hline
Vote-outs & 0 \\
\hline
\end{tabular}
\label{tab:social}
\end{table}
Since the total data samples, i.e., the number of blockchain nodes, are 1000 in our experiment, the number of peers can range from 1 to 999. The maximum number of blocks a node can generate and the maximum trustworthiness score that a node can achieve are set to be 50 and 10, respectively. If there is no history of vote-outs, then the value of vote-out parameter is given as 0, and if there is a vote-out history, value 1 is given.

\begin{figure}[h]
\centering
%   \begin{subfigure}[b]{0.5\linewidth}
%   \centering
    \includegraphics[width=0.24\textwidth]{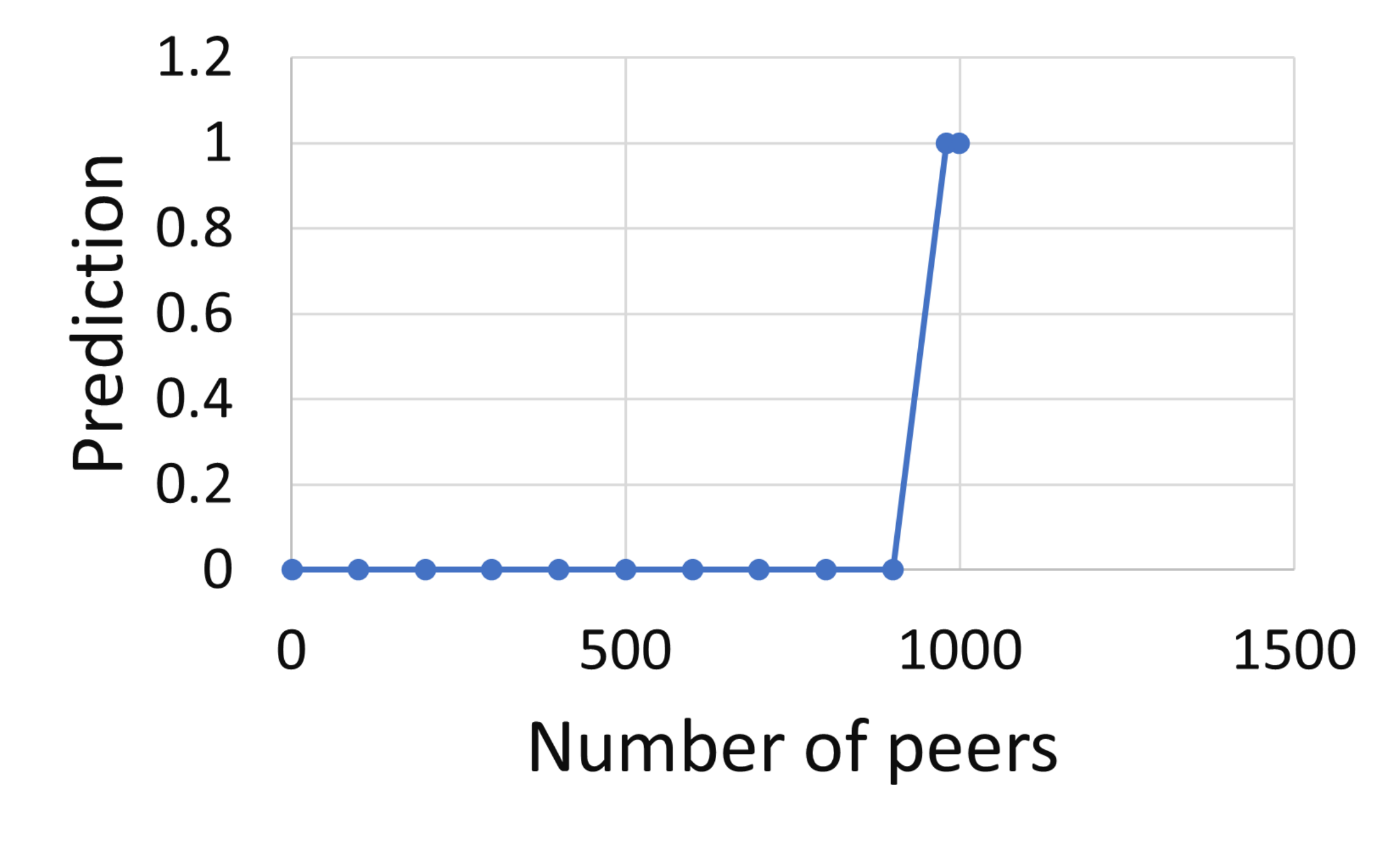} 
%   \end{subfigure}%% 
%   \begin{subfigure}[b]{0.5\linewidth}
%   \hspace{0.5in}
    % \centering
    \includegraphics[width=0.24\textwidth]{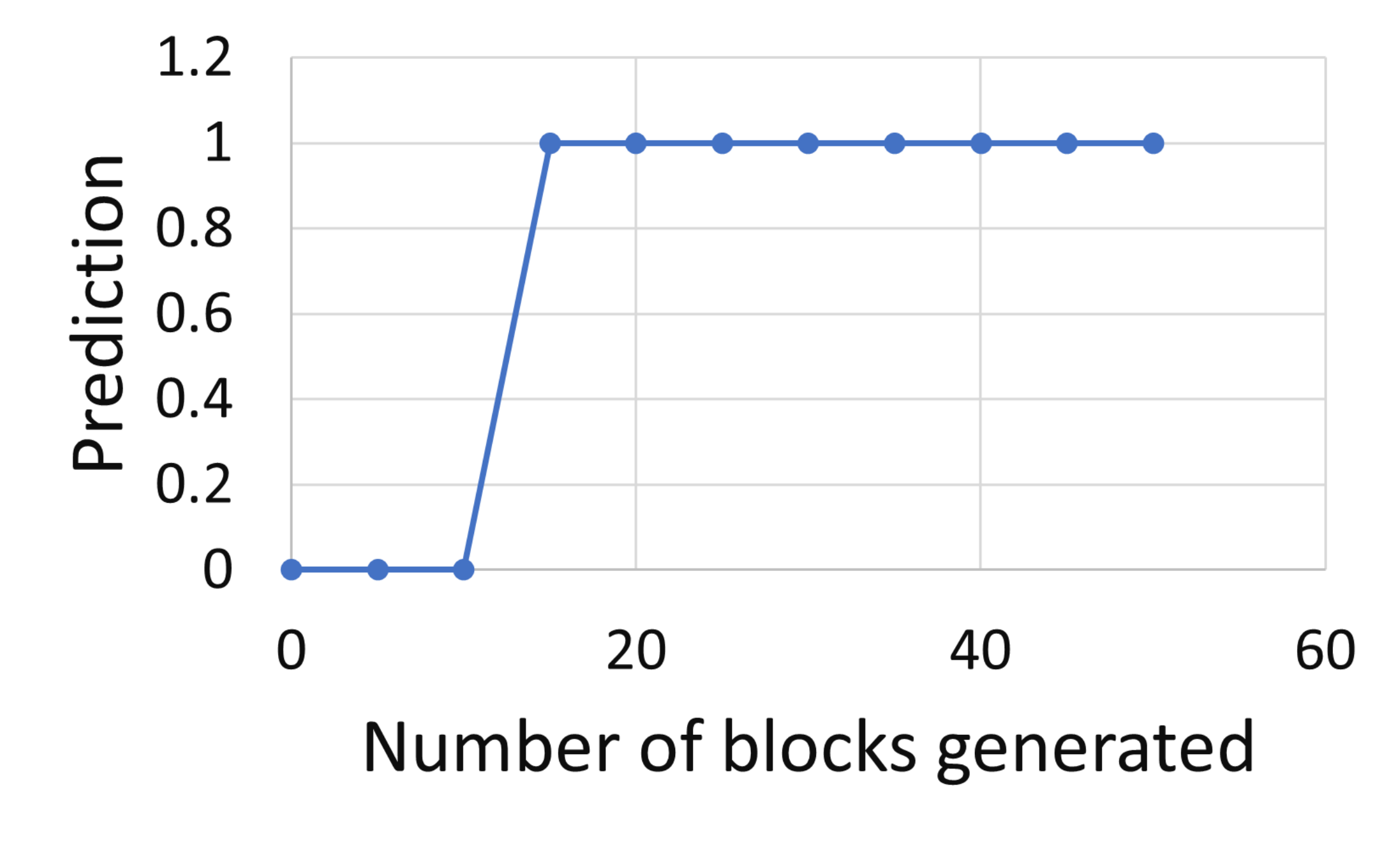} 
%   \end{subfigure}
  \caption{Impacts of the number of peers and the number of blocks generated on leader candidate selection.}
 \label{fig:5} 
\end{figure}

Figure \ref{fig:5} illustrates the impact of  number of peers and number of blocks on the prediction values. On the left side of the figure, the number of peers is changed, keeping all other parameters constant. It can be seen from the figure that when number of peers reaches 980, the prediction becomes 1. This is due to  the fact that when number of peers becomes 980, the total value of the node becomes greater than the threshold value, and   hence the blockchain node is chosen as a leader candidate.  Similarly, for the number of blocks generated parameter, the total value of the node crosses the threshold when the number of blocks generated becomes 15. Hence, at the time at which we run this ML algorithm, if the total number of blocks generated by the node  becomes equal to or greater than 15, it is chosen as a leader candidate. 

\begin{figure}[h]
 \centering
%   \begin{subfigure}[b]{0.5\linewidth}
%   \centering
     \includegraphics[width=0.24\textwidth]{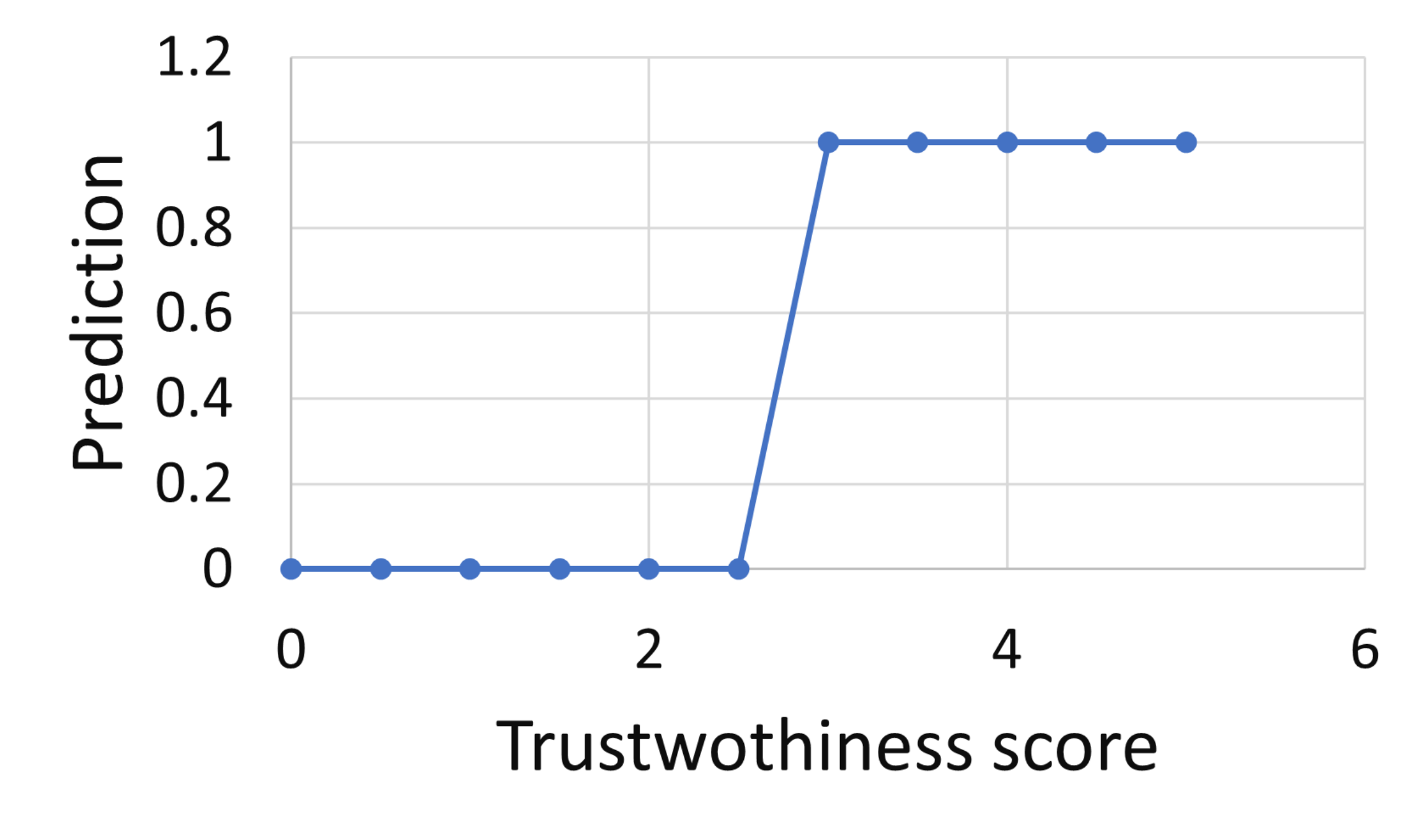} 
%   \end{subfigure}%% 
%   \begin{subfigure}[b]{0.5\linewidth}
%   \hspace{0.5in}
%     \centering
    \includegraphics[width=0.24\textwidth]{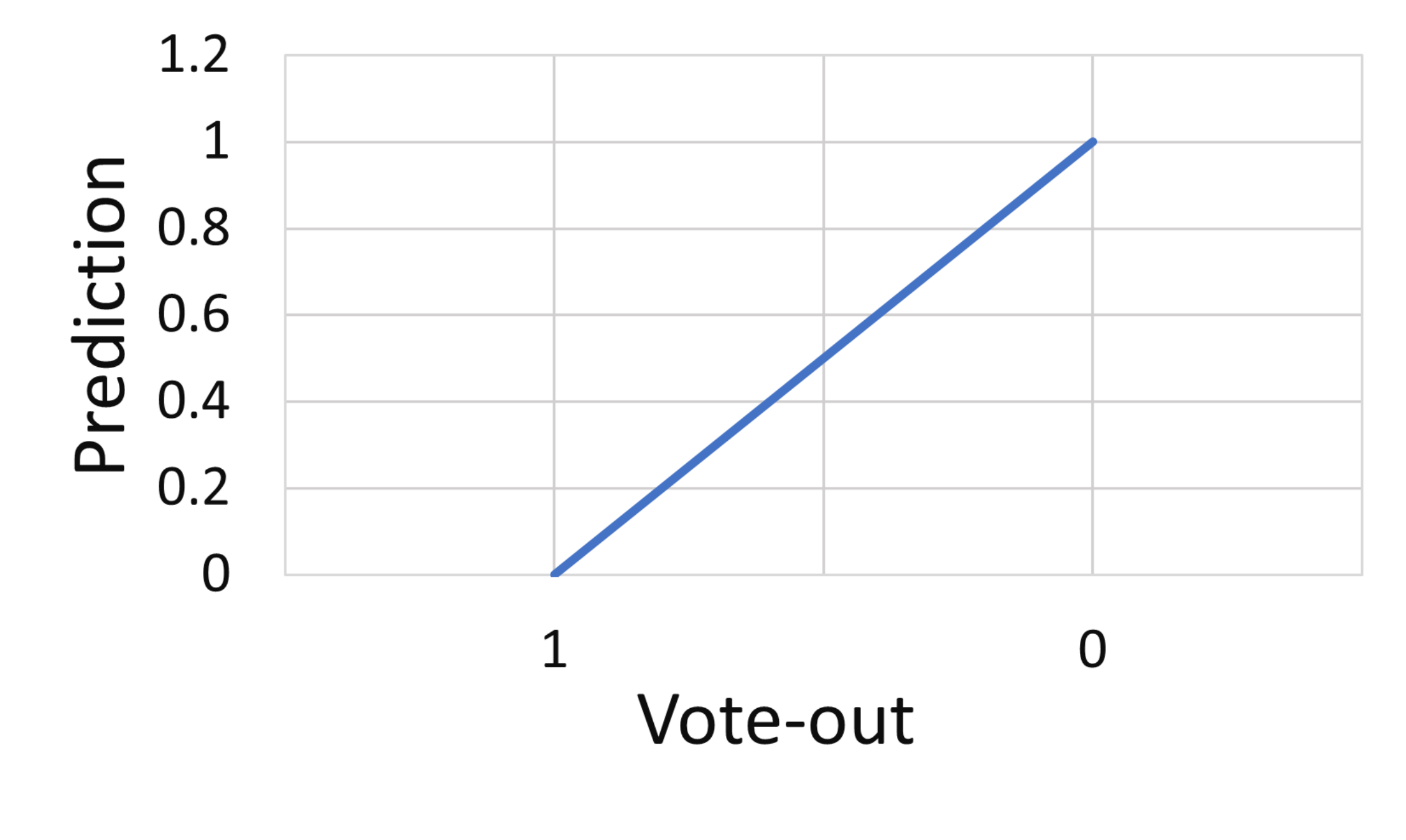} 
%   \end{subfigure}
  \caption{Impacts of the trustworthiness score and the vote-out on leader candidate selection.}
 \label{fig:6} 
\end{figure}

Figure \ref{fig:6} represents the impact of the trustworthiness score and vote-out on the prediction values. In the left side of the figure, trustworthiness score value is changed while keeping the other value constant. We can see that when trust value crosses 3, the prediction changes to 1, since at this point, the total value crosses the threshold value. While analyzing the impact of vote-out, it can be noted that due to higher weights associated with this parameter, vote-out has a significant effect in the prediction values. Hence, if it is 1, then the algorithm will not choose that particular blockchain node as a leader candidate with a higher probability. Since if a node has a history of vote-outs, it could also reflect in its trustworthiness score, and since these two parameters combined have more weights, the probability of that node which has been vote out previously getting chosen as a leader candidate is significantly low. If there is no history of vote-outs, then there is a higher possibility of that node being selected as the leader node.

After leader candidates are selected, modified true random number generator function is invoked which will randomly select a leader from the list and will also provide with the maximum number of blocks that the leader can record in its time.

\subsection{Peer-prediction-based Feedback Collection}
\begin{figure}[hbt]
\centering
\includegraphics[width=0.4\textwidth]{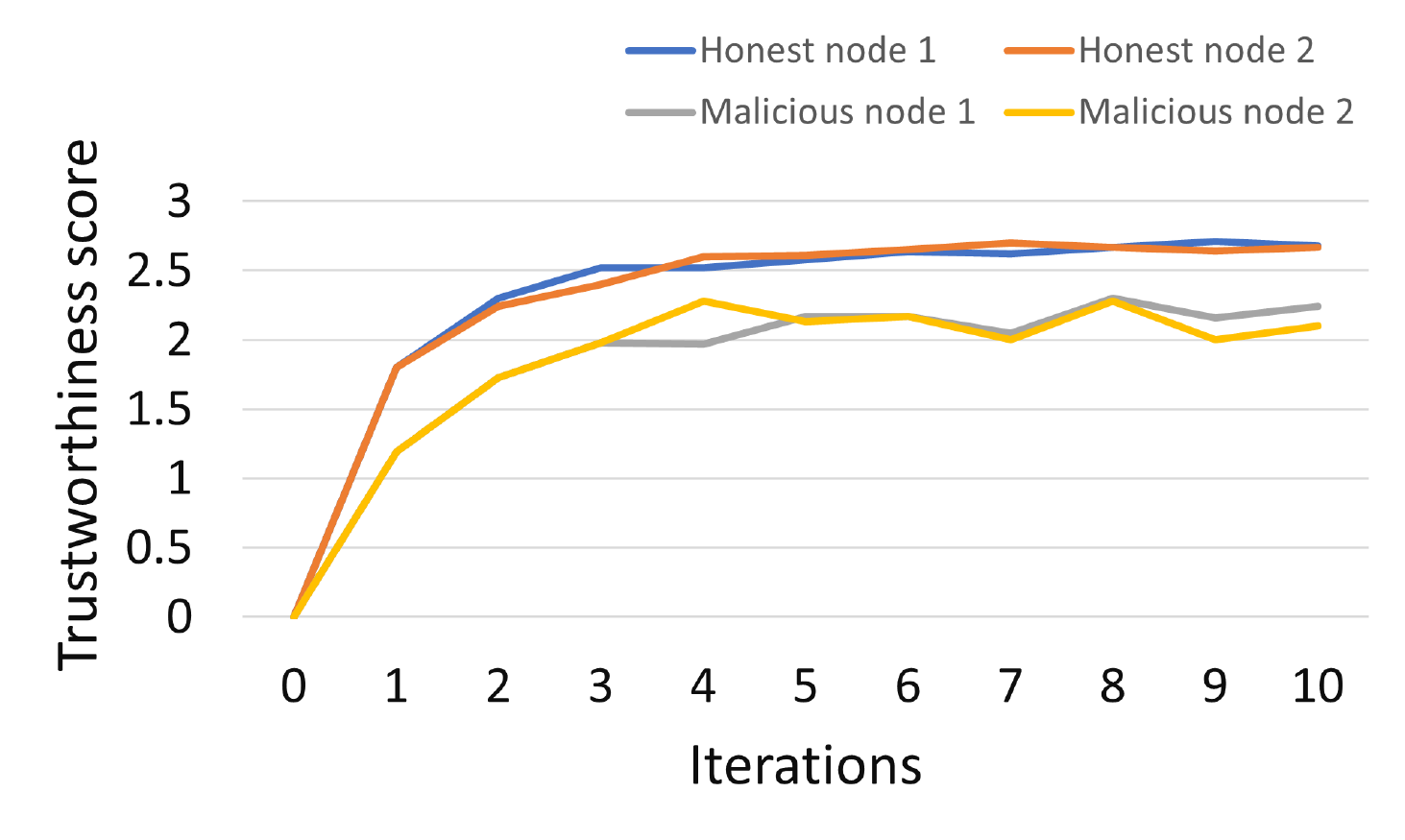}
\caption{Comparison of the trustworthiness values of honest and malicious nodes.}
\label{fig:7}
\end{figure}

In peer-prediction mechanism, the blockchain network is simulated to analyze the significance of trustworthiness value and to study the impact of other parameters on this trustworthiness value. Here, based on the trustworthiness values, the blockchain nodes are classified as honest and malicious nodes. If a blockchain node's trustworthiness value is greater than 50\%, then it is termed as an honest node. Else, the node is treated as malicious.
Figure \ref{fig:7} represents the trustworthiness values of both honest and malicious nodes over 10 iterations. It can be seen that, Honest node 1 and Honest node 2  have  higher trustworthiness values than their peers,i.e., Malicious node 1 and Malicious node 2. The trustworthiness scores of malicious nodes have some irregularities, which can be seen from their graphs. This is because most malicious nodes in the network will try to portray themselves as legitimate nodes;  initially, to avoid suspicion, they will send honest  feedback before sending dishonest ones to confuse its peers. Hence, these malicious nodes have an increase in trustworthiness value which then goes down when these nodes send a dishonest feedback, and this cycle continues. Hence, the graphs of malicious nodes shows irregularities rather than a uniform decreasing trend. 

\begin{figure}[hbt]
\centering
\includegraphics[width=0.4\textwidth]{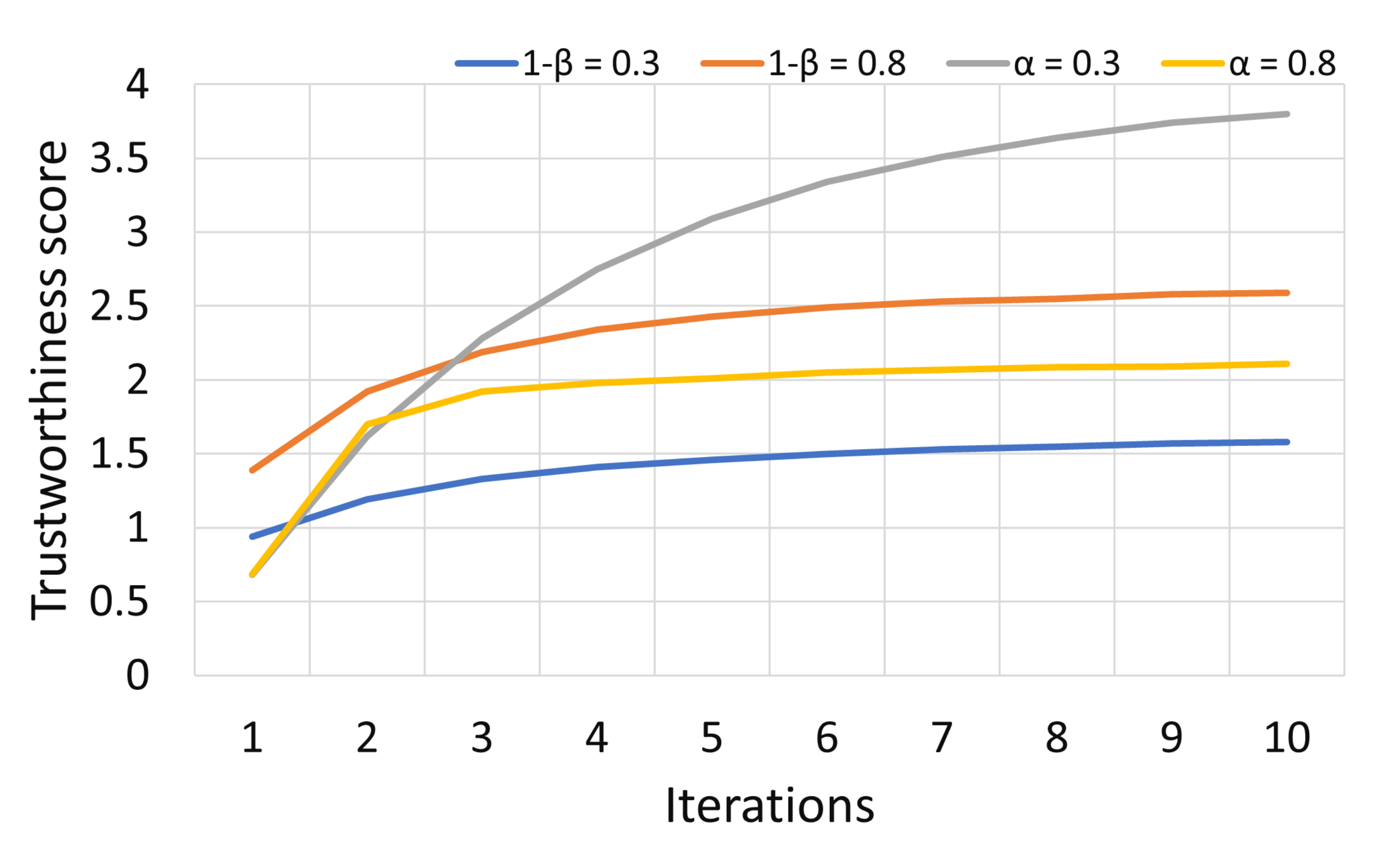}
\caption{Impacts of the promptness index ($1-\beta$) and $\alpha$ (history weight) on the trustworthiness value.}
\label{fig:8}
\end{figure}

Figure \ref{fig:8} represents the impact of promptness index and $\alpha$ (history weight) on the trustworthiness values of the followers. Here, when the promptness index ($1-\beta$)  is changed from  0.3 to 0.8, $\alpha$ value is kept constant at 0.5. Similarly, when $\alpha$ value is changed, the promptness index is kept constant at 0.8.  Since the trustworthiness value is directly proportional to the promptness index ($1-\beta$), the increase in  promptness index leads to the trustworthiness score also increasing. The impact of $\alpha$ on the trustworthiness values of the followers is well-understood by the variables, previous trustworthiness value ($\hat{T}$), and current score value ($R_{q}$) of the followers given in equation (\ref{trust}). Since a follower's $\hat{T}$ is mostly larger than the follower's $R_{q}$ (ranges from 0 to 1), with increase in $\alpha$ value, the trustworthiness score decreases.

%\begin{figure}[h]
 %\centering
%   \begin{subfigure}[b]{0.5\linewidth}
%   \centering
%     \includegraphics[width=0.23\textwidth]{1c.pdf} 
%   \end{subfigure}%% 
%   \begin{subfigure}[b]{0.5\linewidth}
%   \hspace{0.5in}
%     \centering
 %   \includegraphics[width=0.23\textwidth]{1l.pdf} 
%   \end{subfigure}
 % \caption{The effect of parameters: the latency index and $\alpha$ on the trustworthiness value.}
% \label{fig:8} 
%\end{figure}

\section{Conclusion}
In this paper, we have proposed a novel, machine learning-based, blockchain consensus protocol, with transaction prioritization feature for smart city applications. Transaction prioritization feature was implemented to help in recording  transactions with critical information promptly without delay. We designed an algorithm employing a ML algorithm with a random function for an efficient leader election process. The persuasive simulation results show the accuracy and hence efficiency of the employed LightGBM algorithm.  We also proposed a novel dynamic block creation policy for block generation. Further, a peer prediction-based mechanism was designed to encourage honest feedback from the followers for the evaluation of leader's work, and the effect of different parameters on the trustworthiness value was studied. Also, the accumulation of trustworthiness value over a period of time was studied to differentiate the behavior of honest and malicious nodes.

\bibliographystyle{IEEEtran}
\bibliography{ref}

\end{document}